\documentclass[12pt]{amsart}

\usepackage[margin=2cm]{geometry} 
\usepackage{amsmath}
\usepackage{color}
\usepackage{graphicx}
\usepackage{float}

\usepackage{amsmath}
\usepackage{amssymb}

\usepackage{hyperref}
\makeatletter
\def\@secnumfont{\bfseries}
\makeatletter

\makeatletter
\renewcommand\subsection{\@startsection {subsection}{1}{\z@}%
{-2.5ex \@plus -1ex \@minus -.2ex}%
{0ex \@plus.2ex}%
{\normalfont}}
\makeatother

\numberwithin{equation}{section}

\title{Forecasting the 2022-23 tech layoffs using epidemiological models}
\author{Richard Vale}
\email{richardvale@d24h.hk}
\address{Laboratory of Data Discovery for Health, Units 1201-1206, 1223 \& 1225, 12/F, Building 19W, 19 Science Park West Avenue, Hong Kong Science Park, Pak Shek Kok, New Territories, Hong Kong.}
\date{\today}

\begin{document}
\begin{abstract}Many large and small companies in the tech and startup sector have been laying off an unusually high number of workers in 2022 and 2023. We are interested in predicting when this period of layoffs might end, without resorting to economic forecasts. We observe that a sample of layoffs up to March 31, 2023 follow the pattern of noisy observations from an SIR (Susceptible-Infectious-Removed) model. A model is fitted to the data using an analytical solution to the SIR model obtained by Kr\"{o}ger and Schlickeiser \cite{KS20}. From the fitted model we estimate that the number of weekly layoffs will return to normal levels around the end of 2023.
\end{abstract} 
\maketitle

\section{Introduction}
\subsection{}
Many tech workers have been laid off in the period 2022-23 \cite{Sucher} \cite{Hossen}. Self-reported data on tech layoffs is collected by a variety of online aggregators, including \texttt{layoffstracker.com}, \texttt{trueup.io} and \texttt{layoffs.fyi}. The aim of the present paper is to estimate when the ongoing period of unusually high layoffs might end.

\subsection{}
Since the decision of a company to lay off workers tends to be dictated by economic circumstances, the most natural approach to forecasting future layoffs is to use economic forecasts. However, economic forecasts are notoriously unreliable \cite{O14} and it is therefore interesting to look for an approach which does not rely on economic forecasts.

\subsection{}
In the present work, we take the approach that layoffs are mostly a social phenomenon. In a December 2022 interview \cite{StanfordNews}, Professor Jeffrey Pfeffer suggests that the ongoing tech layoffs are a form of ``social contagion" with little or no basis in economic reality. The decision of a company to lay off workers is influenced by the knowledge that other companies are also laying off workers. We therefore investigate whether layoffs can be modelled as a self-sustaining epidemic which is not influenced by external factors.

\subsection{}
The structure of this paper is as follows. In Section \ref{Data} we describe a data set of layoffs. In Section \ref{Model} we explain how the data can be modelled as a set of noisy observations from an epidemiological model. In Section \ref{Results} we explain how the model was fitted to the data and show the results of the analysis. The results are analysed in Section \ref{Discussion}.

\section{Data}\label{Data}
\subsection{}
We obtained a set of data on layoffs which is freely available at Roger Lee's website \texttt{layoffs.fyi}. The data covers 2440 reported layoff events between 11 March 2020 and (at the time of extraction) 30 March 2023. For many companies, the number of people laid off and/or the percentage of the company laid off is given. However, for 15\% of companies, neither statistic is available.

\subsection{}
\begin{figure}[H]
\centering
\includegraphics[scale=0.6]{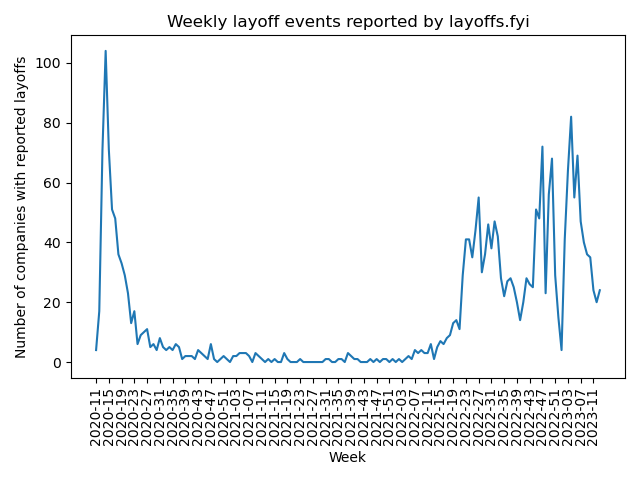}
\caption{Weekly layoffs}
\label{data_all}
\end{figure}

Because we envisage an epidemic of layoffs spreading from company to company, but not from worker to worker, we choose to focus on the number of companies with layoffs, rather than the number of workers laid off. There is a very strong weekly pattern in the number of layoffs. Most layoffs occur on Tuesday, Wednesday and Thursday. We therefore aggregate the layoffs by week in order to avoid seasonality issues. Figure \ref{data_all} shows the weekly data.

\subsection{}
Figure \ref{data_all} shows that, after a peak during the pandemic, reported layoffs returned to a very low level for the whole of 2021. The number of reported layoffs began to climb again in 2022 and has not yet returned to its 2021 level.

\section{Model}\label{Model}
\subsection{}
We choose to model layoffs as an epidemic starting in Week 1, 2022. (Another reasonable choice might be Week 8, 2022, corresponding to the Russian invasion of Ukraine on February 24, 2022.)

\subsection{}
We assume that there is some fixed number $N$ of companies which might lay workers off and where the layoffs might be reported. These $N$ companies form the initial Susceptible population. At the start time, some of the Susceptible population is Infectious. Infectious companies are those which will have layoffs. As time goes by, more and more of the Susceptible population become Infectious. After being Infectious for some time, a company moves to the Removed state and cannot return to either of the previous states. Thus, we neglect the possibility of multiple rounds of layoffs, as these tend to occur only in the biggest companies.

\subsection{}
We assume that layoffs occur at the moment when a company moves from Susceptible to Infectious. Thus, the number of layoffs at time $t$ is the instantaneous increase in the Infectious population at time $t$.

\subsection{}
We choose the dynamics of the model following the SIR model of Kermack and McEndrick \cite{KM}.  This is one of the most basic epidemiological models. Let $S(t)$, $I(t)$ and $R(t)$ denote the Susceptible, Infectious and Removed populations at time $t$. Then we assume that the following equations hold for some $\gamma, \delta > 0$.

\begin{align*}
\frac{dS(t)}{dt} &= - \delta I(t) S(t) / N\\
\frac{dI(t)}{dt} &= \delta I(t) S(t) / N - \gamma I(t)\\
\frac{dR(t)}{dt} &= \gamma I(t)
\end{align*}

\subsection{}
Under the SIR model, the number of layoffs at time $t$ is the instantaneous increase in $I(t)$.
\begin{equation}
\ell(t) = \delta I(t) S(t) / N
\end{equation}
An approximate analytic expression for $j(t) = \ell(t)/N$ (the proportion of the $N$ companies which have layoffs at time $t$) is given in \cite{KS20}, Equations 33, 42, 54, 58, 59, 60, 61, 62, 66, 71, 73, 80, 81, 82, 83. It has a Gaussian shape in the middle, with exponential lower and upper tails. Since we want to estimate the duration of the epidemic, we are particularly interested in the upper tail. The expression for $j(t)$ depends only on the proportion $J(0)$ infected at time $0$, and a parameter $k = \gamma/\delta$.

\subsection{}
We could alternatively model $\ell(t)/N$ as $\int_{t-1}^t j(u)du$. This would correspond to observing all layoffs between weeks $t-1$ and $t$ instead of taking a snapshot at week $t$. Because $\int_{t-1}^t j(u)du \approx j(t)$, this alternative choice of $\ell(t)$ makes very little difference to the results below and the final conclusions of Sections \ref{CI} and Section \ref{conclusion} do not change if $j(t)$ is replaced by $\int_{t-1}^t j(u)du$.

\subsection{}
The observed layoffs at time $t$ are assumed to be a random sample from the actual layoffs $\ell(t)$ at time $t$. We assume that there is a probability $p(t)$ that each layoff occuring at time $t$ is reported. Thus, reported layoffs at time $t$ follow a binomial $\mathrm{Bin}(\ell(t), p(t))$ distribution. If $p(t)$ is assumed to be time-independent, then the variance of the simulated data will be much lower than the variance of the observed data, so we must allow $p(t)$ to vary with time. As is standard in Bayesian statistics, we assume that the $p(t)$ are independent draws from a common distribution, which we take to be a $\mathrm{Beta}(\alpha, \beta)$ distribution, where $\alpha$ and $\beta$ are model parameters.

\subsection{}
Under these assumptions, if $x(t)_{t=0}^T$ is the observed number of layoffs, then $x(t)$ follows a beta-binomial distribution, and the likelihood function for the model is
$$\prod_{t=0}^T {Nj(t) \choose x(t)} \frac{B(x(t) + \alpha, Nj(t) - x(t) + \beta)}{B(\alpha, \beta)}$$
where $B$ is the beta function. The model has five parameters: $J(0)$, $k$, $N$, $\alpha$, $\beta$. The parameters $J(0)$ and $k$ determine $j(t)$.

\subsection{}
\begin{figure}[H]
\centering
\includegraphics[scale=0.6]{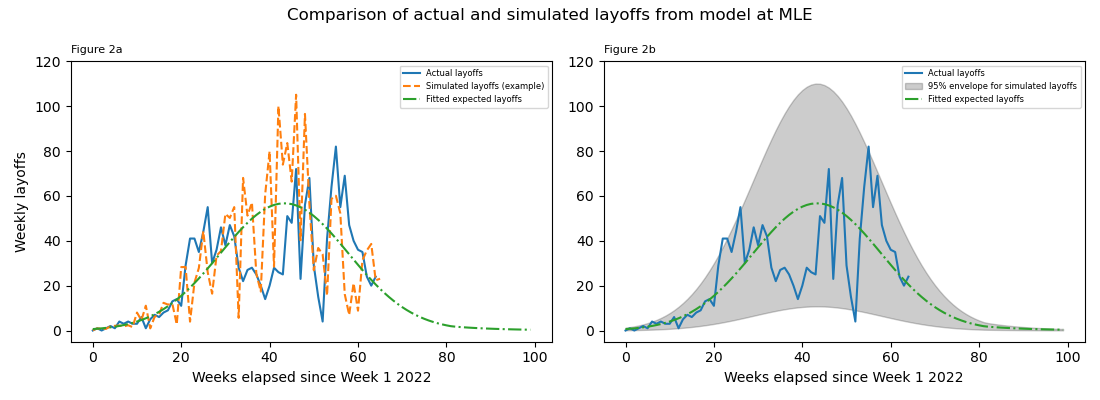}
\caption{Comparison of data and fitted model}
\label{fig2}
\end{figure}

The maximum likelihood estimates of the parameters are $J(0) = 0.00376$, $k = 0.899$, $N = 28261$, $\alpha = 2.15$, $\beta = 2.92$. As a sanity check, we generate some data from the model with these parameters and compare it to the observed data. An example is shown in Figure \ref{fig2}. The simulated layoffs (dotted line in Figure 2a) are random because they depend on $p(t)$. They are plotted in Figure 2a to show that data generated by the model qualitatively resembles the observed data. Figure 2b shows an envelope within which 95\% of simulated layoffs fall.

\section{Results}\label{Results}
\subsection{}\label{baselinechoice}
In order to estimate the duration of the epidemic, we need to decide on a definition of when the epidemic has ended. This definition must be based solely on the data, because we are modelling whether people report layoffs to \texttt{layoffs.fyi}, as well as whether layoffs are actually happening. A natural choice of a baseline level of layoffs is the average reported weekly layoffs for 2021, which is $0.846$.

\subsection{}
We therefore define the end of the epidemic to be the first time $t_{\mathrm{end}}$ at which expected layoffs reach $0.846$. The expected layoffs at time $t$ are given by $Nj(t)\alpha/(\alpha + \beta)$ and they can therefore be computed from the parameters of the fitted model.

$$t_{\mathrm{end}} = \mathrm{min}\{ t : Nj(t)\alpha/(\alpha + \beta) < 0.846 \}$$

\subsection{}\label{CI}
\begin{figure}[H]
\centering
\includegraphics[scale=0.7]{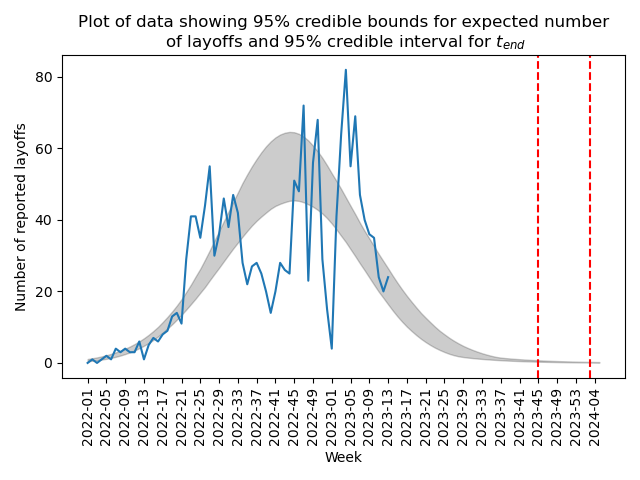}
\caption{Credible interval for $t_{\mathrm{end}}$}
\label{results}
\end{figure}

Bayesian inference was performed with uniform priors on each of the parameters as follows: $J(0) \in [10^{-6}, 1]; k \in [0, 1]; N \in [0, 10^{10}]; \alpha, \beta \in [0, 1000]$. Markov Chain Monte Carlo running for 1000 iterations produced a 95\% credible interval
$$t_{\mathrm{end}} \in [96, 107]$$
which corresponds to the layoffs returning to normal levels between 5 November 2023 and 21 January 2024. The results are shown in Figure \ref{results}.

\section{Discussion}\label{Discussion}
\subsection{}
The choice of model for the epidemic is fairly arbitrary. Underlying the model are the assumptions that the firms mix homogeneously (perhaps not so unreasonable) and also that there is a fixed population of companies which will lay people off, with no new companies, no companies collapsing or being closed, and no repeat rounds of layoffs. We also chose to model layoff events by company, ignoring the severity of layoffs.

\subsection{}
The model proposed here also neglects known sources of non-randomness in the data. For example, the low number of layoffs in late December 2022 probably occurs because of the Christmas/New Year holidays but this is not reflected in the model.

\subsection{}
The choice of the definition of $t_{\mathrm{end}}$ from Section \ref{baselinechoice} is also quite arbitrary and influences the results. It would be conservative to say that we predict that the epidemic of tech layoffs will end around the end of 2023.

\subsection{}\label{conclusion}
\begin{figure}[H]
\centering
\includegraphics[scale=0.6]{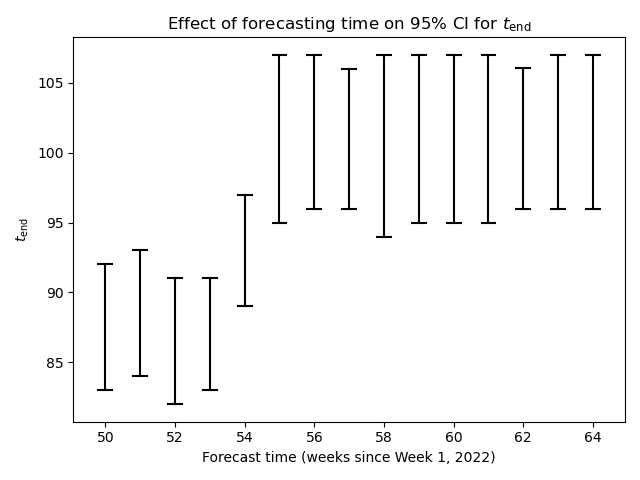}
\caption{Credible intervals for $t_{\mathrm{end}}$ forecast at different times}
\label{sensitivity1}
\end{figure}

In order to check how much the data affects the results, we repeated the analysis using the data only up until week $t$ for $t \in \{50, 51, \ldots, 64\}$. The resulting credible intervals are plotted in Figure \ref{sensitivity1}. It appears that since Week 3 of 2023, the prediction for $t_{\mathrm{end}}$ has been quite stable.

\subsection{}\label{outoftime}
\begin{figure}[H]
\centering
\includegraphics[scale=0.6]{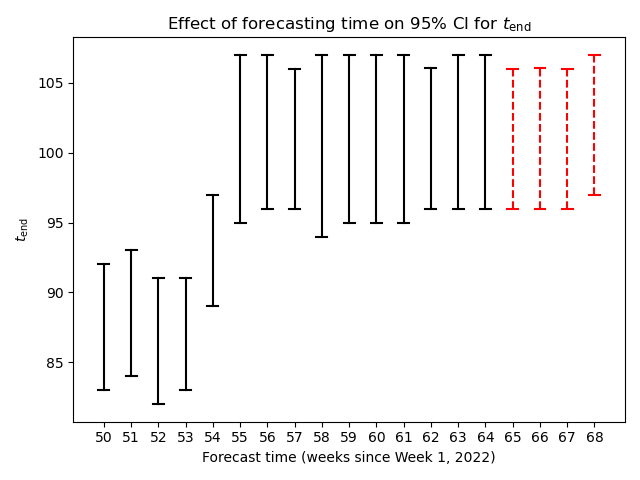}
\caption{Credible intervals for $t_{\mathrm{end}}$ forecast at different times, including four extra weeks of data}
\label{sensitivity2}
\end{figure}

After writing the present paper, we gathered four more weeks of data in order to get an out-of-time sample. The results are shown in Figure \ref{sensitivity2}. Again, it appears that the prediction for $t_{\mathrm{end}}$ has remained stable and it is still reasonable to forecast that the current epidemic of tech layoffs will end around the end of 2023. This prediction is our best guess based on the chosen model and the observed data up to now.

\section{Conclusion}
\subsection{}
Layoffs reported since the start of 2022 by one online layoff tracker seem to follow the pattern of noisy observations from an SIR model of an infectious disease spreading among tech companies. This hints that layoffs may be driven by social factors as well as economic factors. Fitting a model to layoffs suggests that the current period of high layoffs will end around the end of 2023. We will be able to test whether this prediction is correct by the end of 2023.

\section{Acknowledgements}
\subsection{}
We thank Dr. Horace Choi for valuable comments. The paper \cite{Rodrigues} gives a list of some other applications of the SIR model outside epidemiology.

\bibliographystyle{alpha}
\bibliography{epidbib}

\end{document}